\documentclass[fleqn,10pt]{wlscirep}
\usepackage[utf8]{inputenc}
\usepackage[T1]{fontenc}
\usepackage{epstopdf}

\title{Controlled acceleration of GeV electron beams in an all-optical plasma waveguide}

\author[1]{Kosta~Oubrerie}
\author[1]{Adrien~Leblanc}
\author[1]{Olena~Kononenko}
\author[1]{Ronan~Lahaye}
\author[1]{Igor~A.~Andriyash}
\author[1]{Julien~Gautier}
\author[1]{Jean-Philippe~Goddet}
\author[1]{Lorenzo~Martelli}
\author[1]{Amar~Tafzi}
\author[1]{Kim~Ta~Phuoc}
\author[1,2]{Slava~Smartsev}
\author[1,*]{Cedric~Thaury}
\affil[1]{LOA, CNRS, École Polytechnique, ENSTA Paris, Institut Polytechnique de Paris, 181 Chemin de la Huni\`ere et des Joncherettes, 91120 Palaiseau, France}
\affil[2]{Department of Physics of Complex Systems, Weizmann Institute of Science, Rehovot 76100, Israel}

\affil[*]{cedric.thaury@ensta-paris.fr}

\begin{document}

\begin{abstract}
\end{abstract}

\twocolumn 
\flushbottom
\maketitle

{\bfseries 
Laser-plasma accelerators produce electric fields of the order of 100~GV/m, more than 1000~times larger than radio-frequency accelerators~\cite{PhysRevLett.43.267}. Thanks to this unique field strength, they appear as a promising path to generate electron beams beyond the TeV, for high-energy physics~\cite{doi:10.1063/1.3099645}. Yet, large electric fields are of little benefit if they are not maintained over a long distance. It is therefore of the utmost importance to guide the ultra-intense laser pulse that drives the accelerator. Reaching very high energies is equally useless if the properties of the electron beam change completely shot to shot.
While present state-of-the-art laser-plasma accelerators can already separately address guiding~\cite{2006NatPh...2..696L} and control~\cite{2006Natur.444..737F} challenges by tweaking the plasma structures, the production of beams combining high quality and high energy is yet to be demonstrated. 
Here we use a new approach for guiding the laser, and combined it with a controlled injection technique to demonstrate the reliable and efficient acceleration of high-quality electron beams up to 1.1~GeV, from a 50~TW-class laser.
}

Particle accelerators are indispensable tools for science and technology that allow scrutinizing, operating, and even creating matter and light. Conventional accelerators provide energy to charged particles by exciting radio-frequency waves in metallic cavities. The magnitude of accelerating field is ultimately limited by the material breakdown. Since 1960s~\cite{Fainberg1960}, an alternative has been sought in plasmas, which being already ionized cannot break, and thus can theoretically sustain arbitrarily strong electric fields. Today's plasma accelerators driven by  high-power lasers have evolved from early concepts~\cite{PhysRevLett.43.267} to devices which are able to operate accelerating fields over three orders of magnitude higher than radio-frequency cavities. Despite tremendous progress made by laser-plasma accelerators from early demonstrations~\cite{1995Natur.377..606M,Malka1596,2004Natur.431..541F,2004Natur.431..535M,2004Natur.431..538G}, the quality of the delivered electron bunches still  does not match that of conventional machines. As for any accelerator, the basic challenges are first, to place the particles precisely into the accelerating field, and second, to let it run for a sufficient time to achieve an efficient energy transfer. 

The basic recipe for building a laser-plasma accelerator is straightforward: it consists in focusing an ultra-high intensity laser pulse in a gas which is turned into a plasma. As the laser propagates, it expels all plasma electrons out of its way, and thus generates in its wake a positively charged cavity~\cite{RevModPhys.81.1229}. The fields in this cavity, also known as wakefields reach values of the order of 100~GV/m. This simple formula gets more complicated when it comes to bringing electrons into the wakefield. In almost all laser-plasma accelerators, injection is based on the transient heating of plasma electrons to provide them with enough energy to be trapped. Several schemes have been developed to achieve this heating and control the injection; they include optical injection~\cite{2006Natur.444..737F}, injection in a steep density gradient~\cite{2013PhRvL.110r5006B}, localized ionization injection\cite{2011PhRvL.107d5001P,2011PhRvL.107c5001L}, or a combination of these methods~\cite{2015NatSR...516310T}. They allow to produce electron bunches with a relative energy spread at the percent level for 200-300~MeV electrons~\cite{PhysRevLett.102.164801,PhysRevLett.126.104801}, normalized transverse emittance as small as 0.1mm.mrad~\cite{RevModPhys.90.035002}, and a stability in charge and energy of a few percent, limited by that of the laser~\cite{PhysRevLett.126.104801}. As such, the beam quality now approaches that of conventional accelerators, and rapid advances in laser stability and repetition rate~\cite{eupraxia} should bring laser-plasma accelerator into real competition with them, in the sub-GeV energy range. 

\begin{figure*}[ht!]
\centering
\includegraphics[width=\textwidth]{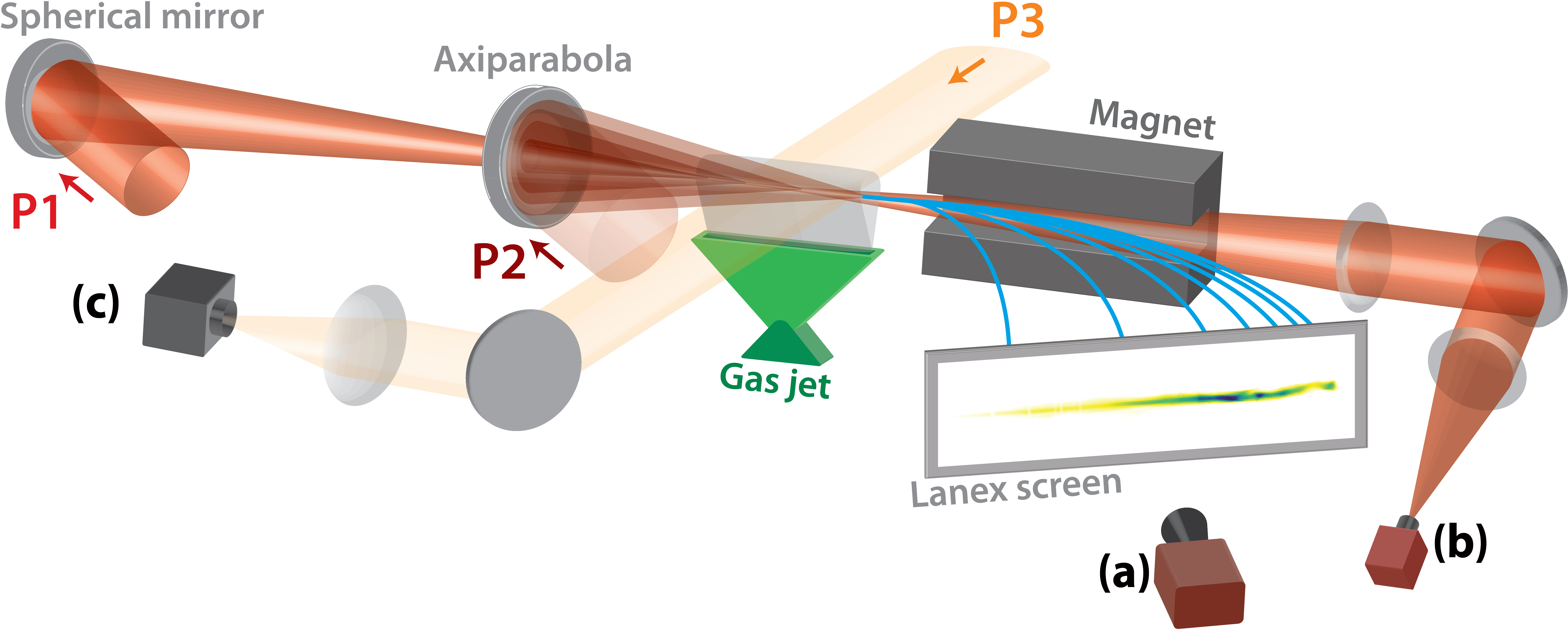}
\caption{ Schematic view of the experiment and main diagnostics. The laser is divided in three beams: P1 the accelerator driver, P2 the pulse that generates the waveguide and P3 a probe beam. The two main beams are focused in a 15 mm long rectangular gas jet. The generation beam P2 is focused in the target, 2~ns before P1, by an f/4 axiparabola. 
The main beam P1 is focused by an f/18 spherical mirror. It accelerates electron beams whose energy is then analyzed by a spectrometer consisting of a dipole magnet, a LANEX scintillating screen and a 16 bits camera (a). The beam P1 is eventually attenuated to image its focal spot after interaction (b). The probe beam P3 crosses  the plasma transversely, just after P1 (c). It is then 
sent to a wavefront sensor  to measure the transverse density profile.} 
\label{fig:schema}
\end{figure*}

The picture is much more complex at  higher energies. The difficulty in that case is to sustain the electric field over a long distance. Achieving the required guiding of  the laser pulse, while preserving a high beam quality, has indeed emerged as one of the most important challenges for laser-plasma acceleration. A breakthrough was achieved in 2008 with the acceleration of an electron beam up to an energy of 1 GeV, in a plasma waveguide~\cite{2006NatPh...2..696L}. The latter consists in a plasma channel whose electron density $n_e$ decreases towards the optical axis. Since the refractive index varies as the opposite of the plasma density, the plasma channel acts as a graded-index optical fiber~\cite{1978ApOpt..17.3990F}, able to guide an intense laser pulse. In the context of laser-plasma acceleration, the only technique demonstrated so far to generate a plasma waveguide is the capillary discharge~\cite{PhysRevLett.89.185003}. This device is a gas-filled capillary to which ends a pulsed high voltage is applied. The discharge ionizes the gas along the capillary axis and produces a hot plasma which then radially expands during a few nanoseconds to generate the waveguide. This technique was successfully used to break several energy records,  from 1~GeV with a 1.6~J laser pulse~\cite{2006NatPh...2..696L}, up to  7.8~GeV with 31~J~\cite{PhysRevLett.122.084801}. It has however a few drawbacks. First, at low plasma density, the channel is not deep enough to effectively guide the laser, meaning that a significant amount of energy can reach the walls of the capillary and damage it. As a consequence, the use of an additional laser pulse to further deepen the plasma channel~\cite{doi:10.1063/1.4793447} is required to accelerate electrons above $\sim 4$~GeV. Moreover, the fact that any leakage of energy out of the guide can damage the capillary raises questions about the possibility of using this device intensively at high repetition rates and large laser energies. Last but not least, this device has never been demonstrated with a controlled injection technique, resulting to the acceleration of unstable and poor quality electron beams.

These  shortcomings has led to the search for new guiding methods.
In particular, it was recently proposed to use a laser pulse to create the plasma channel instead of a discharge~\cite{PhysRevE.97.053203}.
It was in fact a step backward since  the first demonstrated concept of plasma guiding was based on the use of a 100~ps laser pulse which was focused to a line   by an axicon lens  to produce a plasma column from an Ar-gas. The radial expansion of this plasma  then led to the formation of a waveguide, similar to the capillary discharge~\cite{PhysRevLett.71.2409}. The innovation with respect to this pioneering experiment lies in the use of a femtosecond laser pulse, which  allows to produce the plasma through optical field ionization, instead of collisional ionization.  While the latter is not effective at the low densities required for high-energy plasma accelerators ($n_e\lesssim 10^{18}$ cm$^{-3}$), the  efficiency of field ionization does not depend on the gas density. Different implementations of this laser-generated waveguide were proposed and successfully used to transport laser pulses of relativistic intensity ($I \gtrsim 10^{19}$~W.cm$^{-2}$) on a cm length~\cite{Smartsev:19}
, and weaker laser pulses at densities as low as $5\times 10^{16}$~cm$^{-3}$ on a meter scale length~\cite{PhysRevLett.125.074801,PhysRevE.102.053201}.
It is therefore perfectly suited for laser-plasma acceleration, which requires to keep a laser focused at a relativistic intensity in a plasma with density up to few $10^{18}$~cm$^{-3}$. It is also particularly adapted to be used with controlled injection as it leaves a large freedom to shape the plasma density profile, use various gases or multiple laser beams.

Here, we use this laser-generated plasma waveguide to guide an intense laser pulse over 15~mm and show the efficient generation of 1.1~GeV electron beams, from a 1.7~J / 30~fs laser pulse. The versatility of this approach is demonstrated using two different injection methods, ionization injection and density transition injection, leading either to broadband spectra with a high total charge, or electron beams with narrow spectra and a higher charge density at the GeV level.

\begin{figure}[ht!]
\centering
\includegraphics[width=\linewidth]{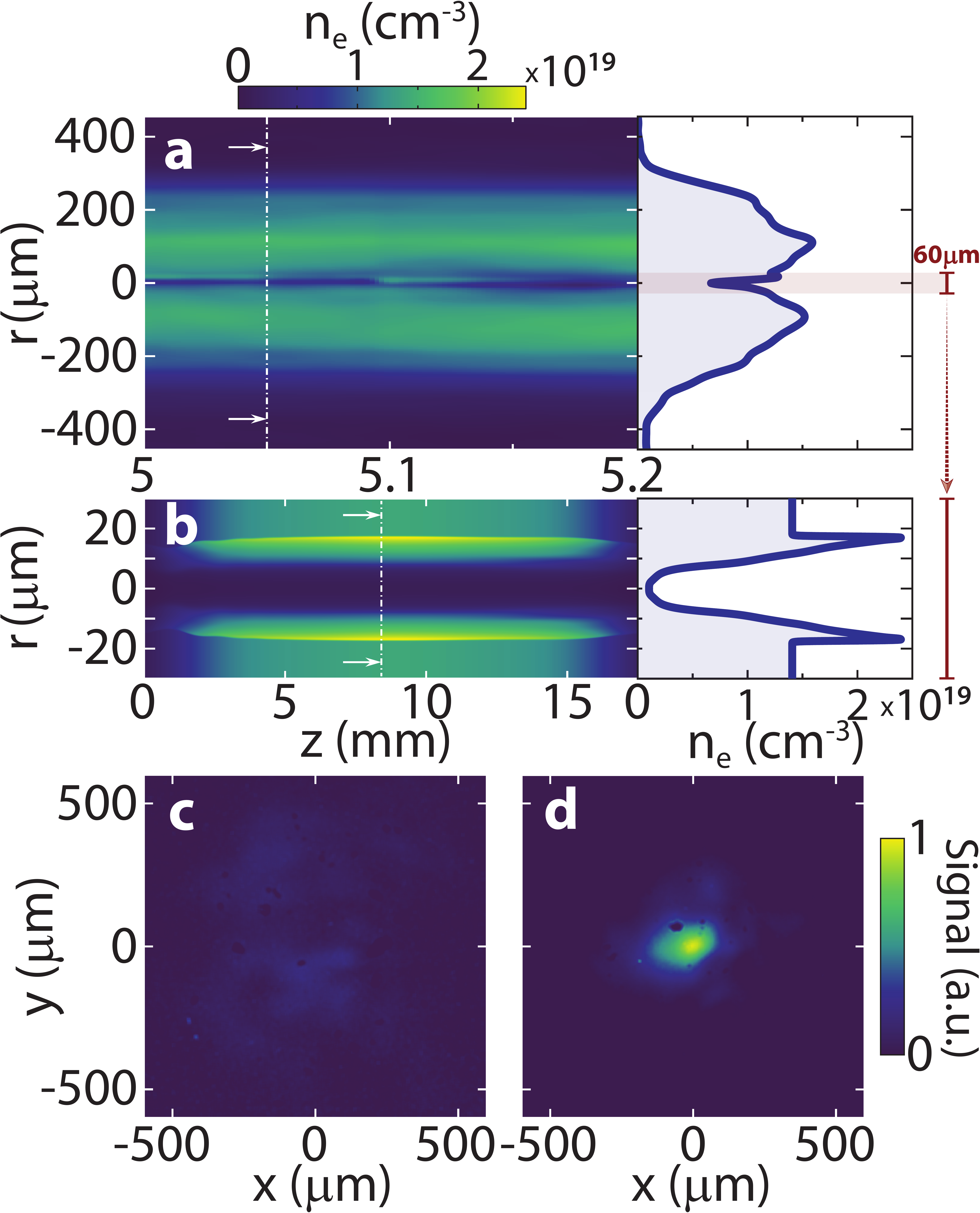}
\caption{Plasma channel and guiding. \textbf{a}, Measured density profile. The density in the middle of the waveguide is overestimated because of limited resolution.\textbf{b}, Simulated plasma channel (see Methods).  \textbf{c-d}, Laser focal spot, measured after attenuation of the full energy beam at the exit of the plasma, without (c) or with (d) guiding.  }
\label{fig2}
\end{figure}

The experimental setup is shown in Fig.~\ref{fig:schema}. A 1.5~mJ, 30~fs laser pulse (P2) is focused  by an axiparabola~\cite{Smartsev:19} into a supersonic gas jet to generate a 15~mm long plasma filament. The plasma then expands radially to generate after 2~ns a plasma waveguide. Figure~\ref{fig2}a shows a typical density profile measured after the formation of the waveguide. The central part of this profile  was simulated using a hydrodynamic code  (see Methods). From these simulations, we  estimate that the  guide has a full diameter of $\sim 30$~$\mu$m, and an axial density of $(1.4\pm 0.3)\times10^{18}$~cm$^{-3}$ (see Fig~\ref{fig2}b).
The main beam (P1) is focused at the entrance of this waveguide, and is effectively guided over 15 mm. This guiding is illustrated in Fig.~\ref{fig2}c-d by focal spots measured at the exit of the plasma, with or without the waveguide (\emph{i.e.} with or without the P2 laser pulse). 

\begin{figure}[ht!]
\centering
\includegraphics[width=\linewidth]{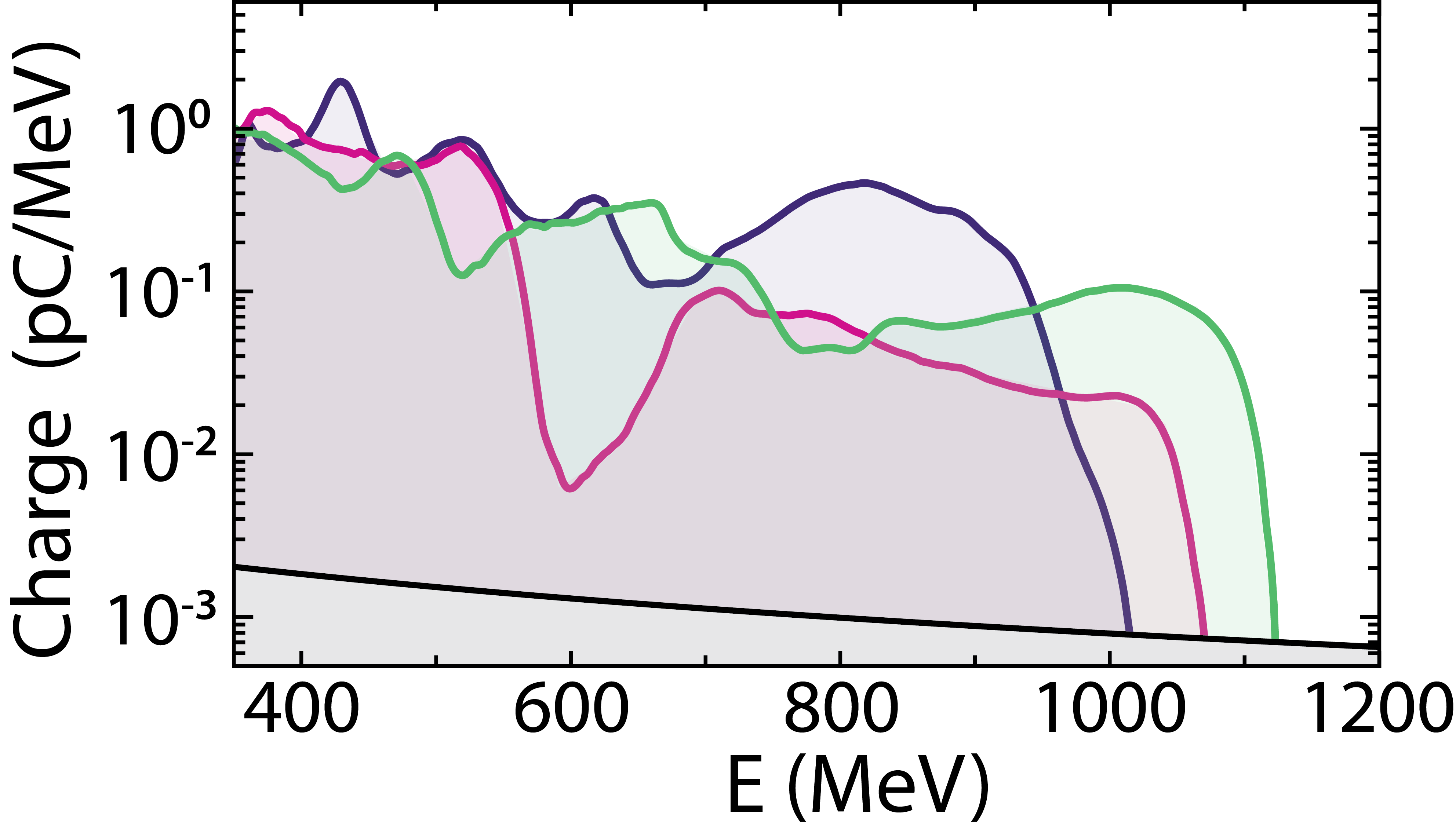}
\caption{Three consecutive spectra resulting from ionization injection. }
\label{fig3}
\end{figure}

Firstly, the setup was used with a gas mixture target so as to trap electrons into the accelerating field through ionization injection~\cite{doi:10.1063/1.2179194}. This injection mechanism, which is used here for its simplicity, generally leads to the generation of electron beams with a broad energy distribution and  a high total charge. 
An example of 3 consecutive spectra obtained in this regime 
is displayed in Fig~\ref{fig3}. As expected, spectra are quasi-continuous with a maximum energy of about 1.1~GeV. The uncertainty on the energy at 1 GeV is of 44~MeV  (see Methods).  The results are consistent with Lu's model~\cite{PhysRevSTAB.10.061301}, which estimates the energy gain and acceleration length in an ideal case, and predicts an acceleration length of $15\pm 3$ mm and an energy gain of $1.3\pm 0.3$~GeV for an electron density $n_e=(1.4\pm0.3)\times10^{18}$ cm$^{-3}$.
The total charge above 350~MeV exceeds 50~pC, so that about 2.2\% of the laser energy in the laser focal spot was transferred to electrons above 350~MeV. Around 70\% of the shots show electrons above 600~MeV; the absence of electrons above this energy for some shots is correlated with a poor guiding of the laser energy, and attributed to pointing fluctuations between the two laser beams (see Methods), which prevents an effective coupling of the main beam into the waveguide. This issue could be solved by using  dynamic correction of the laser pointing~\cite{2011RScI...82c3102G}. 

\begin{figure*}[ht!]
\centering
\includegraphics[width=\textwidth]{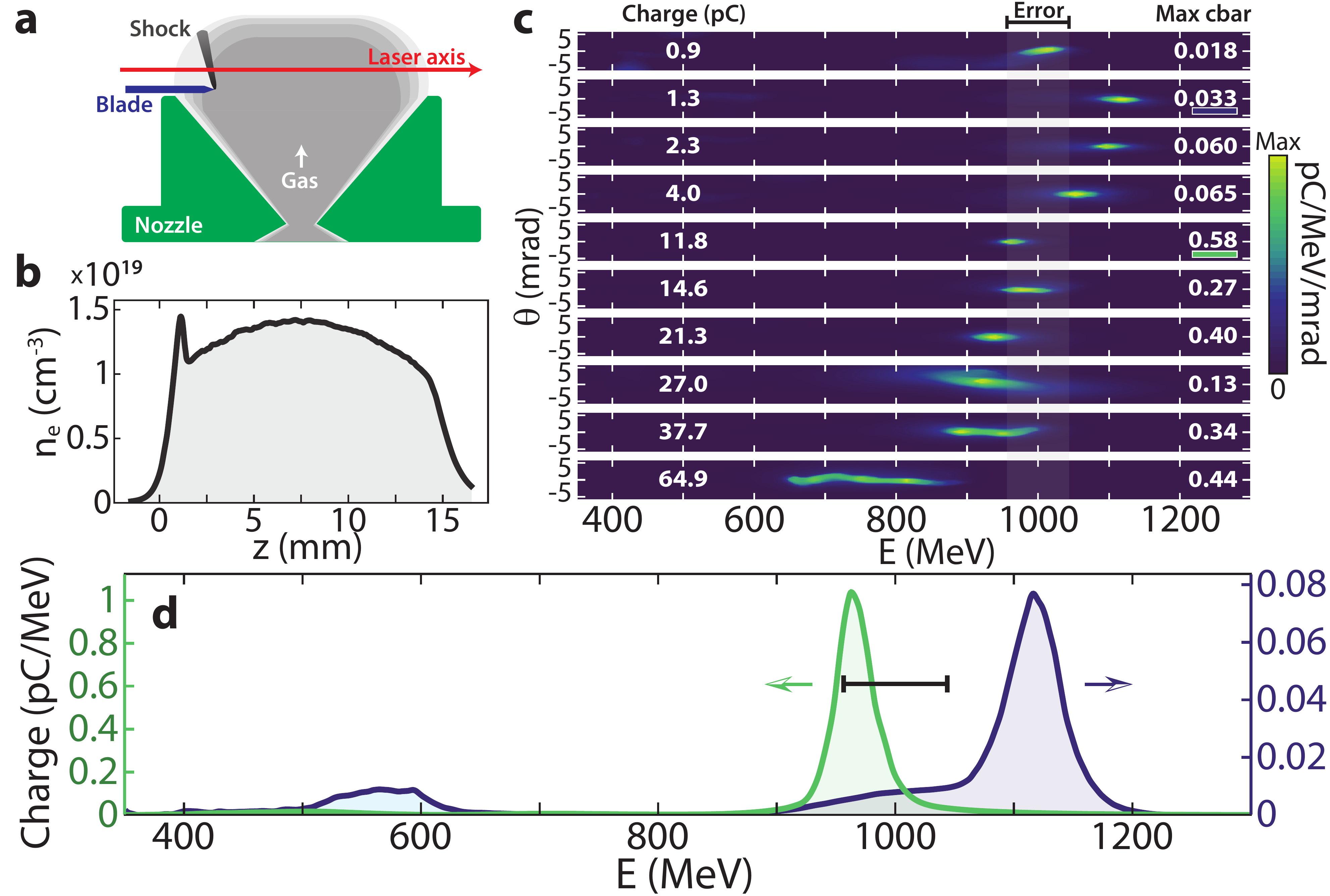}
\caption{Acceleration of high-quality electron beams with density transition injection.  \textbf{a}, Schematic view of the gas target. 
\textbf{b}, Density profile obtained from a fluid simulation for our experimental parameters (see Methods for detailed information). \textbf{c}, Ten angularly resolved electron spectra sorted by charge (charge in the peak for $d^2N/dEd\theta > 0.1(d^2N/dEd\theta)_{\mathrm{max}}$). \textbf{d}, Two examples of angularly-integrated spectra  corresponding to spectra marked by blue and green lines in (c). The black segment indicates  the uncertainty on the energy due to fluctuations of beam pointing.}
\label{fig4}
\end{figure*}

This first experiment demonstrates that an intense laser can efficiently drive a wakefield and then trap and accelerate an electron beam in a laser-generated plasma waveguide. One of the main advantages of this 
approach is that the plasma density can be shaped without affecting the guiding efficiency. Therefore it can be combined with density transition injection to finely control  the trapping position of electrons into the accelerator and thus the final beam energy. This injection strategy requires a sharp density down-ramp in the region of the target where the injection is desired. This down-ramp induces a sharp increase in the length of the accelerator cavity which leads to the injection of electrons from the back of the cavity into the accelerating field~\cite{PhysRevE.69.026409}. In practice, we obtained this density transition by obstructing the gas flow on one side of the nozzle exit to generate a hydrodynamic shock, as first demonstrated in Ref. \citenum{PhysRevSTAB.13.091301} and illustrated in Fig.~\ref{fig4}a-b.

The improvement of the energy distribution, allowed by the control of the injection is illustrated, in Fig.~\ref{fig4}c by a set of 10 spectra, sorted by increasing charge.  These spectra were selected from a series of 14 consecutive shots, excluding those with a negligible charge which likely results from pointing fluctuations between the two laser beams. As observed, a well-peaked  spectrum is obtained for all successful shots. The conversion efficiency from the laser to the high energy peak is about 1\% for 1~GeV beams and can be as high as 6\% for the most loaded ones.  Figure~\ref{fig4}d shows the angularly integrated spectra of two of these shots. The observed energy spread is  of 4.5~\% for the beam in the series with the highest energy (blue curve) and 3.6~\% for the one with the lowest energy spread (green curve). These spreads are biased by the beam divergence; deconvolution from the divergence measured in the vertical direction leads to energy spreads of 3.7~\% for the blue curve and 2~\% for the green one.
Figure~\ref{fig4}c also exhibits a clear correlation between the  charge of the peak and its energy. This can be attributed to beam loading, or in other words, to the screening of the accelerating field by the beam itself. 
A precise control of this loading can allow to flatten the accelerating field so that the entire electron beam experiences the same field. Such fine control was shown to produce energy spread as low as 2~MeV (full width at half maximum) with 200-300~MeV-class laser-plasma accelerators~\cite{PhysRevLett.102.164801,PhysRevLett.126.104801}.

In summary, we demonstrated for the first time the electron acceleration in a laser-generated plasma waveguide, and showed the generation of electron beams in the GeV range using a 50~TW-class laser pulse.  Moreover,  we showed that it can  be combined with a controlled injection technique to obtain high-quality GeV electron beams. In this proof of concept experiment, the energy spread was measured to be below 4\% for best shots.
The coupling efficiency from the main laser beam into the waveguide varies shot-to-shot because of pointing fluctuations. It leads to significant variations of the trapped charge and to about 30\% of missed shots, which is to our knowledge the lowest percentage of missed shots reported so far for electron acceleration in a plasma waveguide (for both laser-generated waveguides and capillary discharges). Implementation of active or passive pointing stabilization should significantly reduce this number. 

Thanks to optical field ionization and its immunity  to laser damage, our approach should be scalable to the most powerful lasers and higher repetition rates. The scheme seems thus  suitable to produce multi-GeV beams for a free-electron-laser or a collider injector. 
The achieved energy is ultimately limited by dephasing, i.e. the electron beam going faster than the laser pulse in the plasma and eventually exiting the accelerating region of the wakefield. This effect can be mitigated by using a rising density profile to increase the energy by up to a factor of two~\cite{PhysRevA.33.2056,2010PhPl...17f3104R,PhysRevLett.115.155002}. As the efficiency of optical field ionization does not depend on the plasma density, the longitudinal density profile in the waveguide could  be optimized to further boost the energy. 

\section*{Methods}

\subsection*{Laser}
The experiment was conducted at Laboratoire d’Optique Appliqu\'ee on a Ti:Sapphire laser system delivering 70~TW on target at a central wavelength of 810~nm, divided in three 30~fs pulses. One arm, containing an energy of 1.7~J is used for electron acceleration. It was focused into the gas jet with a 1.5-m-focal-length spherical mirror, to a focal spot size of 13.5 $\mu$m (FWHM) to reach an intensity of $\approx 2\times 10^{19}$~W.cm$^{-2}$ (normalized vector potential $a_0\approx 3)$. The encircled energy within the first dark ring was estimated to be 60\%. The pointing stability is of 3~$\mu$rad and 2~$\mu$rad Root Mean Square, corresponding to lateral shifts in the focal plane of 4.4~$\mu$m and 3.1~$\mu$m in the horizontal and  vertical planes respectively. The second arm is used for generating the plasma waveguide. It is attenuated to an energy of 1.46~mJ and focused by a holed axiparabola with a nominal focal length of 200~mm and a focal depth of 20~mm, 2~ns before the main pulse. The focal line of the axiparabola is defined by $f(r)= f_0 + 1/a\times \ln(r/R \times e^{a\delta})$ with $a=1/\delta\times \ln\left(R/r_{hole}\right)$ where $r$, $f_0$, $r_{hole}$, $R$ and $\delta$ are respectively the radial coordinate, the axiparabola nominal focal length, the hole in the center of the axiparabola, the axiparabola radius, and the focal line length. The peak intensity at focus is $\approx5\times10^{15}$~W.cm$^{-2}$, and the focal spot diameter at first zero decreases from 15.5~$\mu$m down to 12~$\mu$m at the end of the focal line. The focal spots of both beams were optimized using two independent deformable mirrors. A third low-energy beam was used to probe the plasma.

\subsection*{Target}
For the ionization injection experiment, the gas was a mixture of 99\% Hydrogen with 1\% of Nitrogen. The backing pressure was 40 bars, and the measured electron density is about $1.4\times10^{19}$ cm$^{-3}$. 
For the density-transition injection experiment, a pure hydrogen gas was used. The backing pressure was  40 bars, and the measured electron density in the plateau region of the unperturbed plasma of approximately $1.4\times10^{19}$ cm$^{-3}$. 
In both cases, the density was measured at each shot, using a probe beam to transversely image the target. The phase introduced by the plasma was measured using a wavefront sensor ($160\times 120$ px$^2$ resolution), and the corresponding density profile reconstructed, assuming cylindrical symmetry around the laser axis and using Abel inversion.
It was mounted on a translation stage to be able to reconstruct the full target.  
In addition, the gas flow produced by the nozzle was simulated using the commercial code Ansys FLUENT.

\subsection*{Simulation of the plasma channel formation}
For more details on the channel features, its creation was simulated numerically. 
The field of the axiparabola beam was computed at multiple positions along the focal line, and  the electron temperature and ionization state were calculated, at each position, considering  collisionless above-threshold ionization (ATI) heating. Finally, the plasma channel radial expansion was modeled using a plasma hydrodynamic simulation considering the obtained electron temperature, ionization state and gas density.

We considered an ideal axiparabola  with a  $200$~mm focal distance, a 30~mm-long focal line,  a $38.1$~mm radius and a central hole of ${R_\text{hole}=8}$~mm. The laser beam was initialized on the axiparabola surface with a top-hat spot of ${R_\text{laser}=27.5}$~mm, and a duration of 28~fs. The laser energy was considered to be 0.5~mJ in order to account for the non-ideal features of the laser beam and the axiparabola.  The optical propagation was modeled using the Axiprop library~\cite{axiprop:2020}. More details on such simulations can be found in Ref.~ \citenum{Oubrerie:arXiv2021}. At the different radii and positions along the focal line, the ionization was calculated via ADK model, and the resulting electron energies were calculated from the laser field potential at the moment of ionization~\cite{Rae:PRA1992}. This simplified approach was cross-checked against a more complete particle-in-cell modeling and has shown a good agreement.

For the simulation of plasma expansion, we used the parallel multidimensional Eulerian hydrocode FRONT~\cite{glazyrin:JEPTL2016}. This code uses Riemann solvers for hyperbolic equations and has a number of physical models implemented as modules. Here, we considered separately electron and ion fluids, and computed the electron thermal conductivity and atomic ionization kinetics.

\subsection*{Electron spectrometer}
Electrons are dispersed by a 0.85~T, $400\times80$~mm$^2$ U magnet on a 365~mm-long  Kodak Lanex Regular screen that is imaged by a 16-bit CCD Andor camera. An interference filter at 546 nm  is placed in front of the camera to minimize parasite light. Absolutely calibrated radioactive Tritium light sources are attached to the scintillating screen to provide charge calibrations~\cite{doi:10.1063/1.5041755}. The spectrometer energy and divergence resolution are  0.8\% and 0.4~mrad at 1~GeV. A 500~$\mu$m wide slit is placed between the gas jet and the magnet at 20~cm from the gas jet exit. The angular width of the slit introduces an uncertainty on the electron energy of 44~MeV at 1~GeV. 

\section*{Data availability}
All relevant data  are available from the authors on request.

\section*{Code availability}
The Axiprop code can be found on Github~\cite{axiprop:2020}. Other codes used to generate the results that are reported in the paper are available from the authors on request.

\bibliography{biblio}

\section*{Acknowledgments}
This project has received funding from the European Union’s
Horizon 2020 Research and Innovation programme under grant agreements no. 101004730 LASERLAB-Europe, 730871 ARIES, and 871124 iFAST. We also acknowledge support from the French Agence Nationale de la Recherche (ANR) under reference ANR-19-TERC-0001-01 (TGV project). Authors are grateful to Semyon Glazyrin of VNIIA (Moscow, Russia) for providing access to the FRONT code and advising on its usage.

\section*{Competing interests}
C.T. and S.S. have filed a patent application (no. EP18305810.6) on axiparabolas and their use to generate a plasma waveguide.

\section*{Author contributions statement}

C.T. and KO. designed the experiment. It was prepared and ran by K.O., A.L., O.K., R.L and C.T with support from J.G., L.M and K.T.P.. K.O analyzed  guiding data, density measurements and electron spectra with a support from O.K. for charge calibration. R.L. analyzed laser intensity data.  I.A. provided a numerical support, notably running hydrodynamic and PIC simulations. K.O. and S.S designed the axiparabola. K.O designed the nozzle with support from J.G.. C.T. wrote the paper with strong help from I.A., A.L., O.K. K.O., R.L. and K.T.P.. The laser was operated by J.-Ph.G. and A.T.. The overall project was supervised by C.T..

\end{document}